# The Impact of Artificial Intelligence on Enterprise Decision-Making Processes

Ernest Górka[1], Dariusz Baran[2], Gabriela Wojak[3], Michał Ćwiąkała[4], Sebastian Zupok[5], Dariusz Starkowski[6], Dariusz Reśko[7], Oliwia Okrasa[8]

*Abstract:*

***Purpose:*** *This paper examines how artificial intelligence (AI) enhances enterprise decision-making. It explores the influence of AI tools on decision speed, accuracy, and managerial effectiveness. The study also identifies key human and organizational factors affecting implementation success.*

***Design/methodology/approach:*** *A quantitative survey of 92 enterprises from multiple industries was conducted. Data were collected via a structured questionnaire and analyzed using descriptive and correlation methods. The approach enables identification of relationships between AI adoption, decision efficiency, and organizational barriers.*

***Findings:*** *Results show that 93% of companies use AI, mainly in customer service, data analysis, and decision support. AI improves efficiency and data-driven management but faces barriers such as employee resistance and high costs. Understanding AI mechanisms and managing change are critical competencies.*

***Practical recommendations:*** *Companies should strengthen change management, enhance AI literacy, and address employee adaptation issues. Effective leadership and transparent communication can improve acceptance and outcomes. Aligning AI tools with human decision-making will maximize business value.*

***Originality/value:*** *The paper offers one of the first quantitative assessments of AI's role in enterprise decision-making. It highlights the synergy between algorithmic and human* intelligence. The study contributes to understanding how AI reshapes management practices and organizational agility.

---

[1]*Department of Social Sciences and Computer Science Nowy Sącz School of Business - National Louis University, Poland, ORCID: 0009-0006-3293-5670,*
*e-mail:* [ewgorka@wsb-nlu.edu.pl;](mailto:ewgorka@wsb-nlu.edu.pl)
[2]*The same as in 1, ORCID: 0009-0006-8697-5459, e-mail: dkbaran@wsb-nlu.edu.pl;*
[3]*I'M BRAND INSTITUTE Sp. z o.o., ORCID: 0009-0003-2958-365X,*
*e-mail:* [g.wojak@imbrandinstitute.com;](mailto:g.wojak@imbrandinstitute.com)
[4]*University College of Professional Education in Wroclaw, ORCID: 0000-0001-9706-864X, e-mail: michal.cwiakala@wsk.pl;*
[5]*The same as in 1, ORCID: 0000-0002-7969-4644, e-mail:* [szupok@wsb-nlu.edu.pl;](mailto:szupok@wsb-nlu.edu.pl)
[6]*Pomeranian Higher School in Starogard Gdanski, Institute of Management, Economics and Logistics, Poland, ORCID: 0000-0001-9232-7673,*
*e-mail:* [dariusz.starkowski@twojestudia.pl;](mailto:dariusz.starkowski@twojestudia.pl)
[7]*The same as in 1, ORCID: 0000-0003-4129-0502, e-mail:* [dresko@wsb-nlu.edu.pl;](mailto:dresko@wsb-nlu.edu.pl)
[8]*The same as in 3, ORCID: 0009-0001-5067-229X, e-mail:* [o.okrasa@imbrandinstitute.com;](mailto:o.okrasa@imbrandinstitute.com)

## 1. Introduction

In the era of digital transformation, artificial intelligence (AI) has become a cornerstone of strategic management and organizational competitiveness. The ability of AI systems to process large datasets, generate predictions, and support decision-making processes has reshaped how enterprises plan, evaluate, and act in dynamic markets (Davenport and Harris, 2017; Russell and Norvig, 2021; Zupok, 2024; Zupok, 2025). Despite the growing academic and practical interest in AI, the understanding of *how* this technology specifically affects managerial decision-making within enterprises remains limited and fragmented (Przegalińska and Jemielniak, 2019; Bielińska-Dusza, 2022).

Previous research has largely focused on technological or ethical aspects of AI, such as algorithmic performance, automation, and data governance (Kaplan and Haenlein, 2019; Robaczyński, 2022). However, far fewer studies have empirically explored the *organizational and behavioral dimensions* of AI adoption — particularly how managers and employees interact with AI-based tools during strategic decision-making (Florea and Croitoru, 2025; Madanchian *et al.,* 2024).

This represents a significant research gap, as successful AI integration depends not only on technology itself but also on leadership adaptability, communication quality, and organizational culture (Górka *et al.,* 2025).

Furthermore, while many studies confirm AI's potential to increase efficiency and accuracy, less attention has been given to identifying the barriers that prevent its full implementation in real business contexts, especially within small and medium-sized enterprises (Karski, 2023; Webb, 2018). The lack of empirical data on these organizational challenges limits the ability of companies to formulate effective AI adoption strategies.

The originality of this research lies in its empirical examination of the interplay between AI technology, managerial decision-making, and organizational adaptation. By employing a quantitative approach based on survey data from 92 enterprises across industries, this study bridges the gap between theory and practice. It provides new insights into how AI influences decision-making efficiency, the nature of barriers encountered, and the competencies required for effective implementation.

Based on this identified research gap, the study seeks to answer the following research questions: To what extent does the implementation of artificial intelligence improve the efficiency and accuracy of enterprise decision-making processes? What organizational and human factors facilitate or hinder the successful adoption of AI in managerial practice? Which skills and competencies are most critical for integrating AI into strategic decision-making?

By addressing these questions, the study contributes to the growing body of literature on digital management and provides both theoretical and practical implications for enterprises navigating the AI-driven transformation of business processes.

## 2. Literature Review

Artificial intelligence (AI) has become one of the most transformative technologies shaping contemporary management theory and practice. Initially defined as computer systems capable of performing tasks traditionally requiring human intelligence - such as recognition, language processing, learning, and decision-making - AI is now regarded as a strategic asset that enhances organizational agility and competitiveness (Kożuchowska, 2002; Rich and Knight, 1983; McCarthy, 2007).

Over time, the concept evolved from early cognitive approaches focused on replicating human reasoning toward more pragmatic perspectives emphasizing performance optimization and data-driven decision support (Russell and Norvig, 2021; Kaplan and Haenlein, 2019). Today, AI represents an interdisciplinary field integrating computer science, statistics, and management science to enable systems that interpret information, learn autonomously, and make complex decisions in dynamic business environments (Nilsson, 2010).

AI's increasing integration into business processes has redefined decision-making models and managerial practices across industries. By enabling the rapid analysis of vast amounts of data, AI supports strategic decisions that are more evidence-based, timely, and precise than those made solely on human judgment (Gliszczyński *et al.*, 2018; Cieplak *et al.,* 2018; Zupok, 2025). Modern enterprises employ AI to enhance organizational performance through decision support systems, predictive analytics, and process automation, which collectively contribute to greater operational efficiency and strategic responsiveness (Kożuchowska, 2002; Przegalińska and Jemielniak, 2019).

Machine learning - a core subset of AI - plays a crucial role in enabling organizations to derive actionable insights from data. In marketing, supervised learning algorithms classify customers by preferences and behaviors, enabling the design of personalized offers and targeted campaigns (Cieplak *et al.,* 2018).

Deep learning techniques, capable of processing unstructured data such as images or text, further expand the range of managerial applications by supporting brand monitoring, sentiment analysis, and trend forecasting (Przegalińska and Jemielniak, 2019). Unsupervised learning, on the other hand, assists companies in identifying previously overlooked market segments and optimizing pricing strategies, improving the effectiveness of commercial initiatives (Mierzwińska *et al.,* 2018).

AI's predictive capabilities extend beyond marketing into financial management, where algorithms detect anomalies, evaluate risk, and forecast cash flows. Classification models such as decision trees and support vector machines are widely used to identify fraudulent activities and assess creditworthiness (Abu Samara *et al.,* 2024; Duarte *et al.,* 2018).

Regression-based forecasting helps organizations anticipate demand fluctuations, develop accurate budgets, and plan investments (Martins *et al.,* 2018). By automating these analytical tasks, AI not only reduces human error but also enables managers to focus on strategic decisions rather than routine operational activities (Bartol and Tylicki, 2017; Harara *et al.,* 2024).

The influence of AI extends across key business functions. In finance and accounting, automation tools streamline repetitive tasks such as document processing and reporting, reducing costs and improving accuracy (Prędkiewicz and Biegun, 2024; Ziółkowska, 2023). In risk management, real-time anomaly detection enhances organizational resilience and protects corporate reputation by preventing fraud (Bojar *et al.,* 2018; Kłoczko, 2024). Within customer relationship management, predictive analytics enables companies to anticipate customer needs, personalize interactions, and improve loyalty (Pawlicka and Bal, 2021; Krystian and Zaskórski, 2023).

AI is also transforming supply chain management. Predictive algorithms optimize demand forecasting and inventory planning, minimizing the risk of shortages or overstocking (Pawlicka and Bal, 2021; Kulisz *et al.,* 2018). Real-time tracking and route optimization enhance logistics performance, reduce costs, and improve delivery times (Bojar *et al.,* 2018; Skowron *et al.,* 2018).

Furthermore, integrating AI with Internet of Things (IoT) technologies and blockchain ensures end-to-end visibility, security, and traceability throughout the supply chain (Skuza and Lizak, 2023; Duarte *et al.,* 2018). These applications highlight AI's capacity not only to improve efficiency but also to support broader strategic goals, such as sustainability and innovation (Jemielniak and Przegalińska, 2019).

One of AI's most significant contributions lies in enhancing decision-making processes. Automated analytics accelerate information processing and reduce the time required to make strategic choices (Cieplak *et al.*, 2018; Baran *et al.*, 2025). AI systems uncover hidden patterns and trends within complex datasets, providing decision-makers with insights that would be difficult or impossible to identify manually (Martins *et al.*, 2018; Bielińska-Dusza, 2022). Moreover, algorithmic decision-making minimizes the influence of cognitive biases, leading to more objective and consistent outcomes (Balbaa and Abdurashidova, 2024; Kulisz *et al.*, 2018).

Importantly, AI complements rather than replaces human judgment. Managers increasingly rely on hybrid decision-making approaches that combine human intuition with machine-generated insights, resulting in more nuanced and strategically sound choices (Przegalińska and Jemielniak, 2019; Skowron and Kulisz, 2018). This synergy enhances organizational adaptability and fosters a culture of data-driven decision-making, strengthening competitive advantage in volatile markets (Gliszczyński *et al.*, 2018).

Despite its transformative potential, the implementation of AI in management is not without challenges. High costs associated with infrastructure, software, and specialized expertise can limit adoption, particularly among small and medium-sized enterprises (Karski, 2023; Webb, 2018). Integration with legacy IT systems often requires significant investment and organizational change, while data fragmentation and quality issues remain major obstacles to effective AI deployment (Wójcik, 2018). Resistance from employees - driven by fears of job displacement or inadequate training 0 can further slow implementation efforts (Przegalińska and Jemielniak, 2019).

Ethical, legal, and security concerns also shape the adoption landscape. Issues of accountability, bias, privacy, and intellectual property introduce complex regulatory and governance challenges (Robaczyński, 2022; Kumalski, 2022). Algorithmic opacity, particularly in deep learning models, raises questions about transparency and trust (Otto, 2022). Cybersecurity threats require continuous investment in encryption, anomaly detection, and workforce training to protect sensitive data and maintain operational integrity (Kalińska-Kula, 2017; Loftus *et al.*, 2020).

The literature consistently demonstrates that AI enhances organizational decision-making, operational efficiency, and strategic agility across a wide range of business functions (Davenport and Harris, 2017; Kerzner, 2017). However, adoption remains uneven, with many organizations - particularly in emerging markets - struggling to fully exploit AI's potential due to cost, skill, and infrastructure barriers (Skowronek-Mielczarek, 2021). Future research should therefore explore how firms can overcome these challenges and integrate AI more effectively into their strategic decision-making processes. Such studies will be crucial to understanding how AI can evolve from a technological tool into a foundational driver of organizational innovation, resilience, and competitive advantage.

Effective management relies not only on the deployment of advanced technologies but also on the human and relational dimensions of organizational life. Internal communication remains a cornerstone of strategic management, shaping how information is interpreted, decisions are executed, and teams align around shared goals. Empirical evidence shows that organizations with clearer messaging, accessible leadership, structured feedback loops, and the strategic use of digital tools demonstrate higher levels of engagement, inclusion, and operational efficiency (Baran *et al.*, 2025).

Recent studies provide evidence for that synergy. Florea and Croitoru (2025) examine how AI-based tools affect internal communication dynamics and employee performance by improving six key communication dimensions: informing, message reception, feedback, acceptance, persuasion, and reaction. They show that AI support reduces transmission errors and enhances clarity in communication processes, thereby boosting overall performance.

Moreover, emerging research emphasizes that the integration of AI into managerial communication processes fundamentally reshapes how leaders interact with their teams. By leveraging real-time analytics, predictive feedback systems, and adaptive communication platforms, leaders can craft more targeted messages, anticipate employee concerns, and cultivate a more inclusive and responsive communication climate (Madanchian *et al.*, 2024).

Leadership style further amplifies the effectiveness of communication systems and mediates the organizational impact of AI adoption. Research based on Blanchard's situational leadership framework shows that supportive (affiliative) styles — emphasizing empathy, trust-building, and interpersonal cohesion — significantly strengthen morale and team cohesion, but require clear goal-setting and continuous feedback to avoid performance stagnation (Górka *et al.*, 2025).

Such leadership qualities are critical in AI-enabled environments, where employees must adapt to algorithmic decision-support tools and integrate automated insights into their daily workflows. Leaders who can flexibly shift between supportive, instructional, and delegating styles are more successful in guiding teams through technological transitions and embedding AI into core processes (Górka *et al.*, 2025).

Beyond enhancing communication, AI fundamentally transforms leadership paradigms by promoting more adaptive, data-driven, and personalized approaches to managing teams. As Madanchian *et al.* (2024) argue, AI tools allow leaders to predict talent needs, analyze performance patterns in real time, mitigate ethical biases, and customize leadership development to individual employees. This convergence of technology and leadership fosters a management model in which human intuition coexists with algorithmic intelligence, enhancing leaders' capacity to inspire, coordinate, and sustain organizational performance while maintaining ethical and human-centered decision-making.

At the project level, leadership directly influences the success of technology-enabled initiatives. Empirical studies show that constructive feedback, goal clarity, and empowerment — combined with encouragement of team initiative — significantly improve project outcomes and stakeholder satisfaction (Ćwiąkała *et al.,* 2025).

These managerial behaviors complement AI applications that optimize planning, resource allocation, and risk assessment, demonstrating that technology alone is insufficient without human direction. Ultimately, the intersection of AI capabilities with effective communication and adaptive leadership forms a synergistic model of modern management - one in which data-driven insights are translated into action through human judgment, trust, and collaboration.

### 3. Research Methodology and Case Description

The research presented in this article was designed to examine and quantitatively determine the role of artificial intelligence (AI) in corporate decision-making processes, with particular emphasis on practical conditions, barriers, and competencies necessary for its effective implementation. The methodological approach used is based on empirical research standards in the social sciences, with an emphasis on collecting measurable data and systematically analyzing respondents' opinions.

The study used a quantitative strategy to identify relationships between variables and analyze trends in a statistically verifiable manner. Empirical data was obtained using a diagnostic survey method, which allows for the identification of attitudes, opinions, and experiences of a specific research group, while maintaining a high degree of comparability of results. The research tool used was a structured survey questionnaire, developed specifically for the purposes of this study.

The questionnaire was designed to gather information on the level of implementation of AI-based solutions in companies, assess their impact on decision-making efficiency, and identify barriers and challenges accompanying the implementation process.

The design of the survey was preceded by an analysis of the literature on the applications of artificial intelligence in management and decision sciences, which ensured its substantive relevance and methodological consistency.

The research process was conducted in the form of an online survey, which ensured a wide reach and accessibility for respondents from different regions and sectors of the economy. Participation in the study was entirely voluntary, and all participants were informed about the anonymity and scientific nature of the project.

*Ernest Górka, Dariusz Baran, Gabriela Wojak, Michał Ćwiąkała, Sebastian Zupok,*
*Dariusz Starkowski, Dariusz Reśko, Oliwia Okrasa*

The survey involved 92 respondents representing various industries and types of enterprises. The diversity of the research sample made it possible to obtain cross-sectional data on the implementation of artificial intelligence in the decision-making processes of organizations.

The collected data was subjected to quantitative analysis. Basic statistical techniques were used, including the calculation of the percentage share of individual response categories and the presentation of results in graphical form – using bar and pie charts. This method of processing the empirical material allowed for a clear presentation of the results and their subsequent comparison with the theoretical data presented in the literature section.

The research methodology adopted ensured the reliability and repeatability of the results obtained, while enabling an objective presentation of the scale of artificial intelligence use in enterprises and the factors influencing the effectiveness of its implementation.

### 4. Research Results

The research conducted is exploratory in nature and focuses on identifying the factors determining the effectiveness of artificial intelligence implementation in enterprises, as well as on recognizing the barriers and opportunities accompanying this process. Its aim is to gain a deeper understanding of the impact of AI-based solutions on decision-making processes in organizations and to identify factors that facilitate or hinder effective implementation.

The questionnaire consisted of three parts: an introduction (preamble) presenting the purpose of the study, a main part containing substantive questions, and a data sheet enabling the characterization of the sample under study. The structure of the tool allowed us to obtain detailed data necessary to analyze the phenomenon and formulate conclusions about the role and significance of artificial intelligence in the management processes of modern enterprises.

The results presented indicate that the vast majority of the companies surveyed already use artificial intelligence-based solutions. As many as 93% of respondents confirmed the implementation of AI in their operations, while 7% declared that they do not use this type of technology. These results show the current level of artificial intelligence use in the surveyed group of companies, taking into account both companies that have implemented such solutions and those that have not yet done so.

The second pie chart presents data on the areas in which companies use artificial intelligence. Respondents were asked: "In what areas does your company use AI?" The chart shows the distribution of responses in four main categories of application for this technology.

***Table 1.*** *The use of artificial intelligence in businesses*

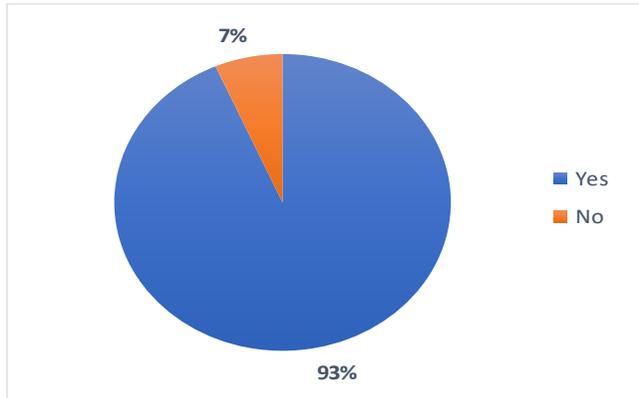

*Source: Own elaboration.*

***Table 2.*** *Areas of application for artificial intelligence in businesses*

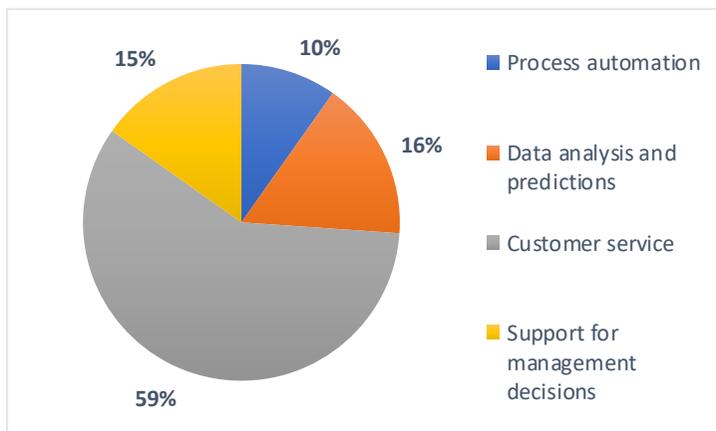

*Source: Own elaboration.*

The analysis shows that artificial intelligence is most often used in customer service – this was indicated by 59% of respondents. This result confirms the importance of AI as a tool that supports communication with customers and improves service quality. Another important area is data analysis and predictions (16%), which highlights the growing role of artificial intelligence in interpreting data, identifying trends, and forecasting future events.

The use of AI in management decision support was declared by 15% of respondents, which indicates the growing importance of this technology in strategic planning and decision-making processes. On the other hand, 10% of respondents indicated process automation, confirming the potential of artificial intelligence in streamlining operational activities and increasing organizational efficiency.

**Table 3.** *The impact of AI on decision-making*

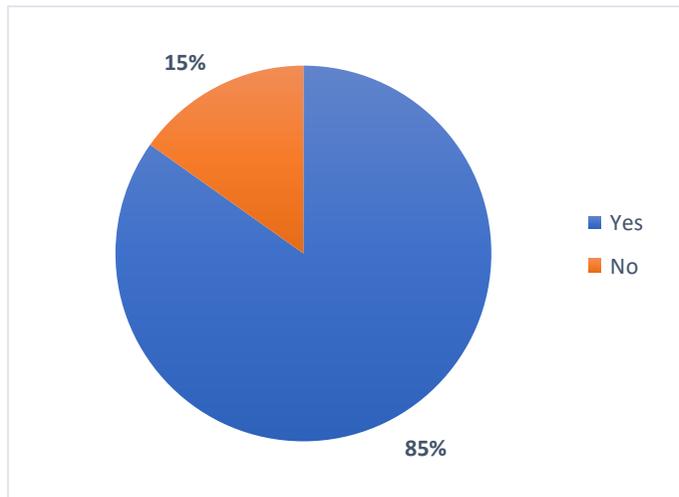

***Source:*** *Own elaboration.*

Respondents participating in the survey were asked to assess the impact of artificial intelligence on the decision-making process in their companies. The vast majority of them (85%) believed that AI-based solutions support this process and contribute to its improvement.

**Table 4.** *AI as an alternative to humans in decision-making*

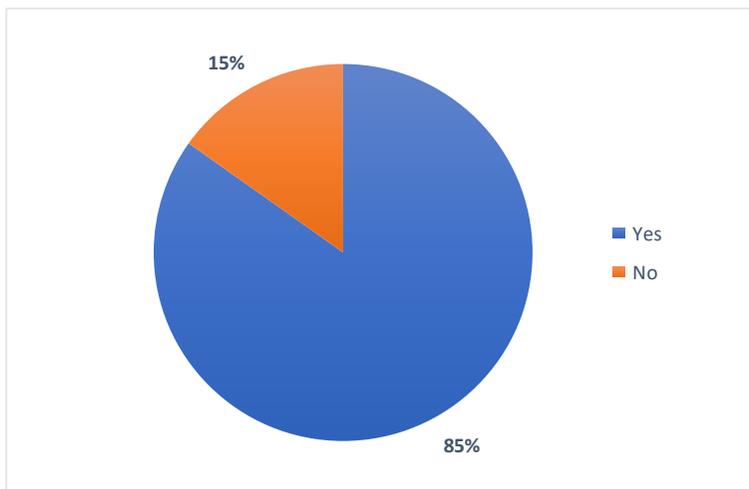

***Source:*** *Own elaboration.*

The respondents also addressed the issue of replacing humans with artificial intelligence in strategic decision-making. Here, too, the prevailing opinion was that this technology has great potential – 85% of respondents indicated that AI is capable of taking over the role of humans in this area.

***Table 5.*** *Does artificial intelligence help increase the effectiveness of decision-making processes?*

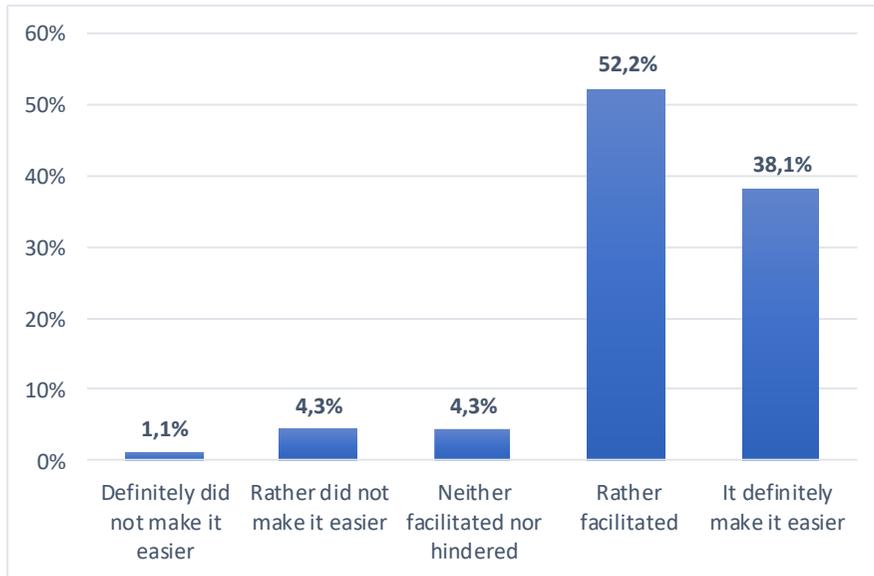

***Source:*** *Own elaboration.*

In addition, survey participants assessed the extent to which the implementation of artificial intelligence has affected the ease of decision-making in their organizations. The bar chart shows the distribution of respondents' answers on a five-point Likert scale, assessing the impact of artificial intelligence implementation on facilitating decision-making.

The response "Definitely did not make it easier" received 1.1%, "Rather did not make it easier" – 4.3%, "Neither facilitated nor hindered" – 4.3%, "Rather facilitated" – 52.2%, and "Definitely make it easier" – 38%. These results indicate a clearly positive trend in the assessment of the impact of artificial intelligence implementation – the majority of respondents considered that the use of AI contributed to facilitating the decision-making process.

The sixth bar chart shows the results of responses to the statement: "Artificial intelligence helps to increase the effectiveness of decision-making processes." The response "I strongly disagree" received 2.2%, "I rather disagree" – 7.6%, "Neutrally" – 15.2%, "I rather agree" – 41.3%, and "I strongly agree" – 33.7%. The data indicates that the majority of respondents agree with the statement that artificial intelligence supports increased efficiency in decision-making processes. Most respondents express a positive belief in the impact of AI on improving decision-making, while only 9.8% of respondents express some degree of disagreement with this thesis.

*Table 6.* *Artificial intelligence helps increase the efficiency of decision-making processes.*

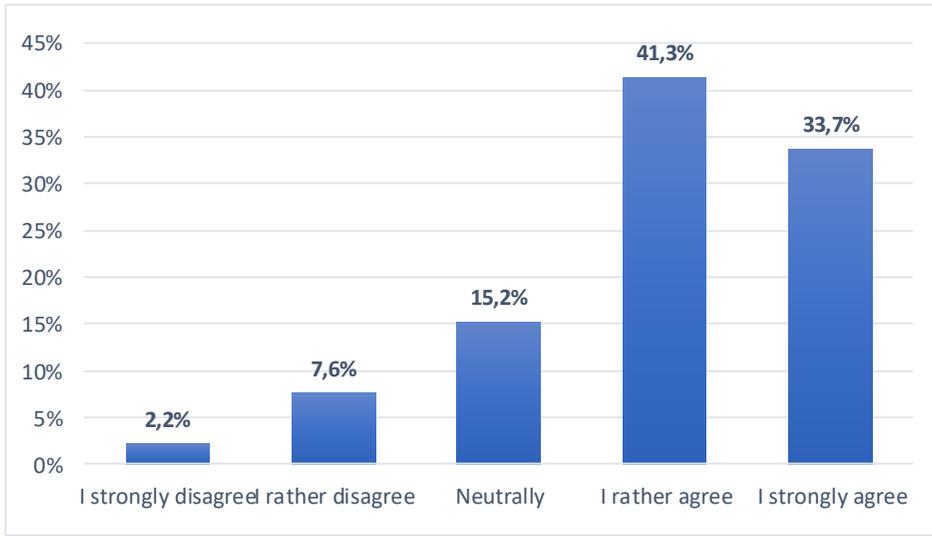

*Source: Own elaboration.*

*Table 7.* *The implementation of AI in a company faces more organizational barriers than technological ones. processes*

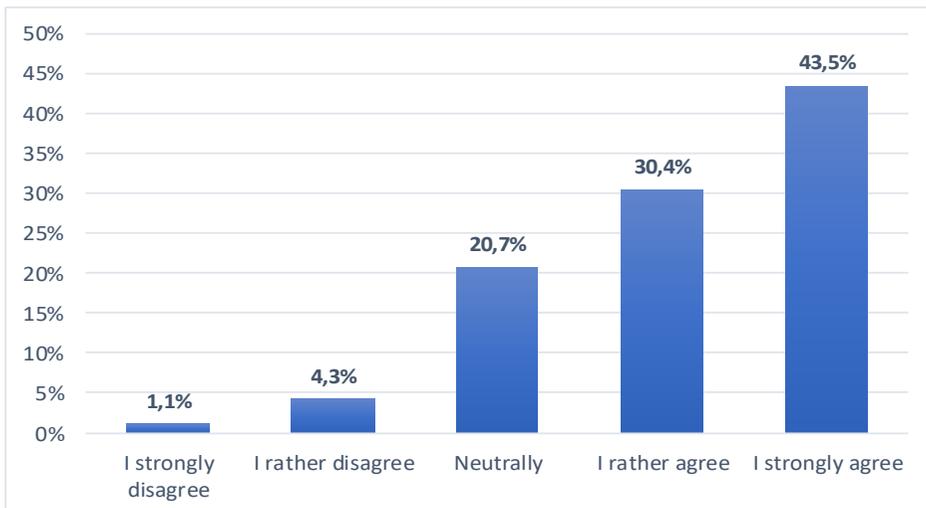

*Source: Own elaboration.*

The next bar chart shows the results of responses to the statement: "The implementation of AI in a company encounters more organizational than technological barriers." The response "I strongly disagree" received 1.1%, "I rather disagree" – 4.3%, "Neutrally" – 20.7%, "I rather agree" – 30.4%, and "I strongly agree" – 43.5%.

The data indicates that the majority of respondents agree with the statement that the implementation of artificial intelligence in companies faces more organizational than technological barriers. The prevailing opinions emphasize that the challenges associated with AI implementation are more often due to organizational factors than technological limitations, while only 5.4% of respondents disagree with this statement.

*Table 8.* Barriers to AI implementation

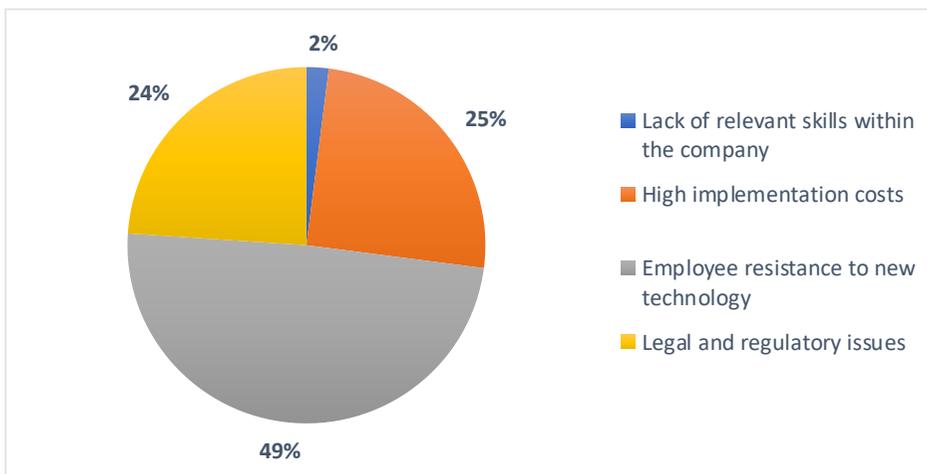

*Source: Own elaboration.*

The next chart shows the results of responses to the question: "What barriers did you encounter when implementing AI?" The diagram shows the percentage share of individual barriers identified among respondents. The most frequently cited problem is employee resistance to new technology, which was mentioned by 49% of respondents.

Another significant barrier is the high cost of implementation, indicated by 25% of respondents, which reflects the financial challenges associated with implementing artificial intelligence. Legal and regulatory issues, highlighted by 24% of survey participants, underscore the ambiguity of regulations governing the use of AI. Finally, the lack of relevant skills within the company, indicated by 2% of respondents, although less common, may also hinder the implementation of artificial intelligence in organizations.

**Table 9.** *Skills in AI implementation*

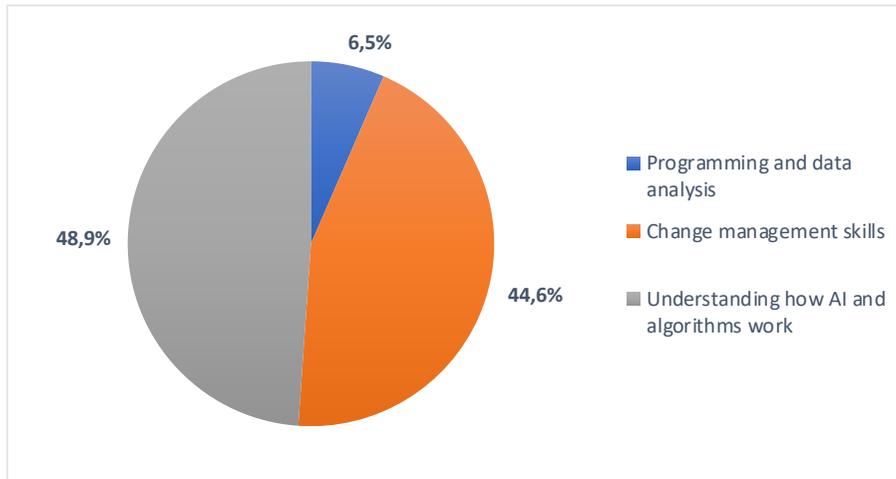

**Source:** *Own elaboration.*

The chart shows the results concerning key skills necessary for the effective implementation of artificial intelligence in enterprises. The analysis of the responses shows that the most important competence is understanding how AI and algorithms work, which was indicated by 48.9% of respondents. This demonstrates the growing importance of technological awareness and the ability to interpret the mechanisms behind artificial intelligence-based solutions.

Slightly fewer, 44.6% of respondents, considered change management to be a key skill – a competency necessary in the process of adapting organizations to new technologies, overcoming employee resistance, and effectively integrating AI into existing structures and processes. The least frequently mentioned, but still important, skill is programming and data analysis (6.5%). This result suggests that respondents place greater emphasis on the practical application of AI-based tools and understanding their impact on the functioning of the company than on the technical aspects of coding itself.

The bar chart shows the results of responses to the question: "Did employees in your company have difficulty adapting to AI tools?" The response "They definitely had no difficulty" received 1.1%, "Rather did not have difficulty" – 0%, "Neutrally" – 4.3%, "Rather, they had difficulties" – 45.7%, and "They definitely had difficulties" – 48.9%.

The results indicate that the vast majority of respondents noticed adaptation difficulties among employees, with almost half assessing them as significant. These data suggest that the implementation of artificial intelligence tools in many companies was associated with challenges in adapting staff to new technologies, which may be due to a lack of experience, fear of change, or the need to acquire new digital skills.

*Table 10. Assessment of the significance of barriers to AI implementation*

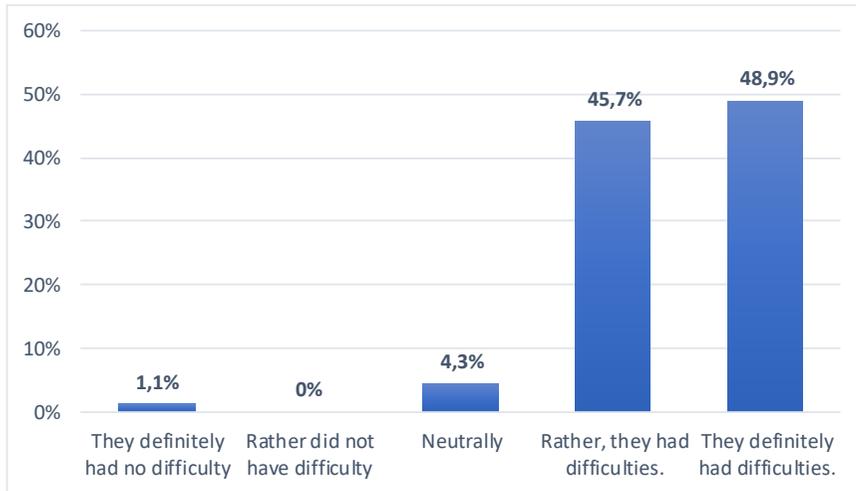

*Source: Own elaboration.*

*Table 11. The problem of employees in AI implementation*

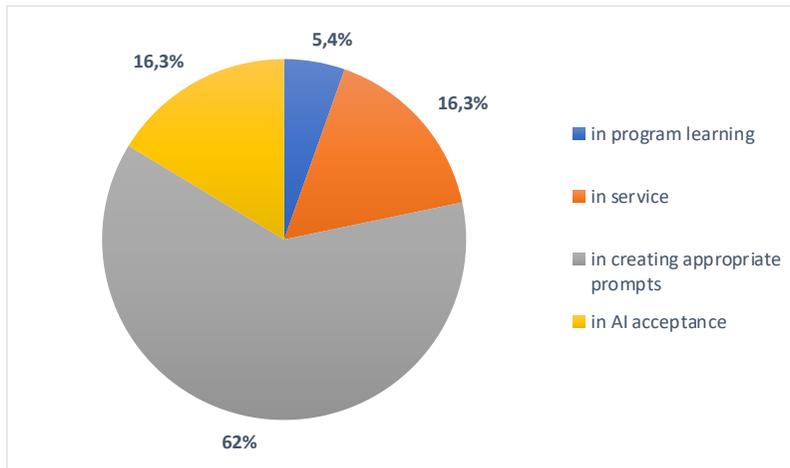

*Source: Own elaboration.*

The pie chart shows the results of responses to the question: "What was the biggest problem for employees?" The diagram shows the percentage share of individual difficulties identified among respondents. The most frequently mentioned challenge was creating appropriate prompts (62%), which highlights the importance of the ability to formulate clear and precise queries when working with artificial intelligence tools, especially generative models.

This was followed by AI acceptance and tool operation, each with 16.3%. These results suggest that some employees may have encountered difficulties due to unintuitive interfaces, complex functions, or a lack of trust in the technology. The least problematic area was learning the programs (5.4%), which may indicate that employees are relatively good at mastering the basic functions of AI tools, although they still need support in using them effectively.

*Table 12. Risks in AI implementation*

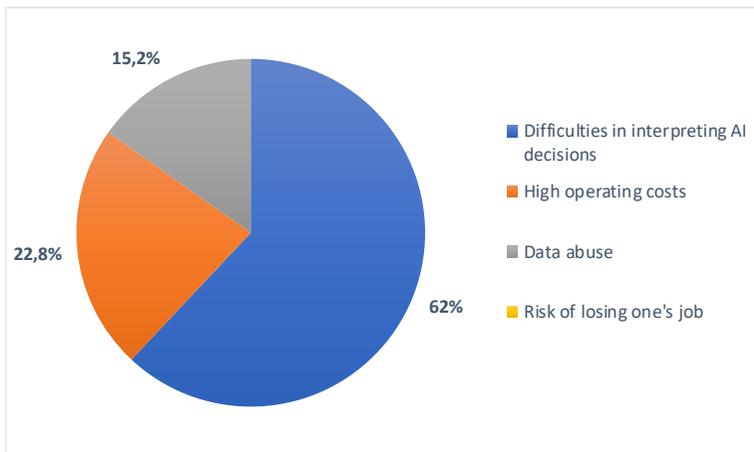

**Source:** *Own elaboration.*

The pie chart shows the results of responses to the question: "What risks do you see in the implementation of AI?" The diagram shows the percentage distribution of respondents' indications regarding the potential risks associated with the implementation of artificial intelligence in enterprises. The most frequently mentioned risk is the difficulty in interpreting decisions made by AI systems (62%), which highlights concerns about the transparency and comprehensibility of algorithms.

In second place were high operating costs (22.8%), indicating the financial aspect of the risk associated with maintaining and developing AI solutions. Another risk is data abuse (15.2%), including issues such as privacy violations, unauthorized access, and unethical use of information.

Importantly, none of the respondents identified the risk of job losses as a significant threat, suggesting that in the companies surveyed, the implementation of artificial intelligence is seen as a support rather than a threat to employment.

Most respondents participating in the survey live in medium-sized cities with up to 150,000 inhabitants (54.3%). This accounts for over half of the sample.

The next largest groups are respondents living in towns with up to 50,000 inhabitants (17.4%) and large cities with over 150,000 inhabitants (also 17.4%). The percentage of respondents living in rural areas is the smallest, at 10.9%.

***Table 13.*** *Place of residence*

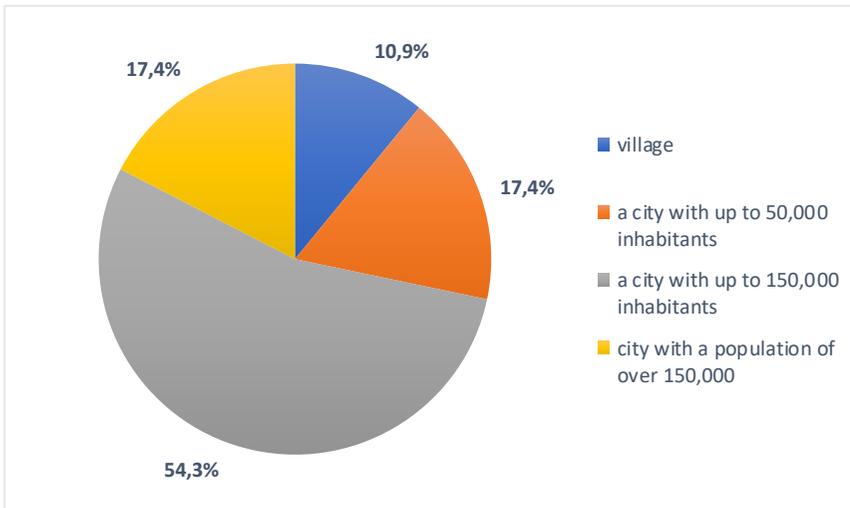

***Source:*** *Own elaboration.*

***Table 14.*** *Respondents' industry*

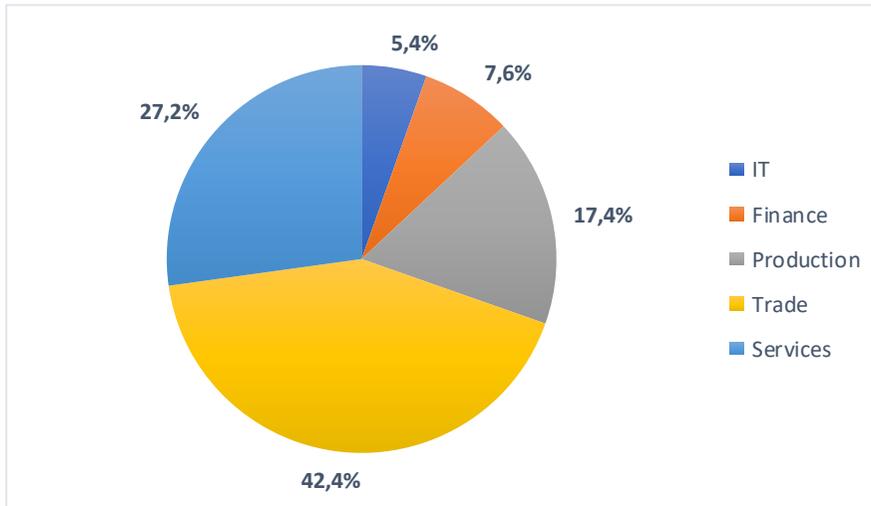

***Source:*** *Own elaboration.*

The most numerous industry represented in the sample is trade (42.4%). This indicates a significant share of companies among the respondents. The second largest group are companies from the service sector (27.2%). Manufacturing (17.4%) is another significant group of respondents. The share of respondents from the IT and finance industries is smaller, but their presence in the survey is also important from the point of view of analyzing the implementation and use of artificial intelligence.

*Table 15.* Correlation matrix between key research variables

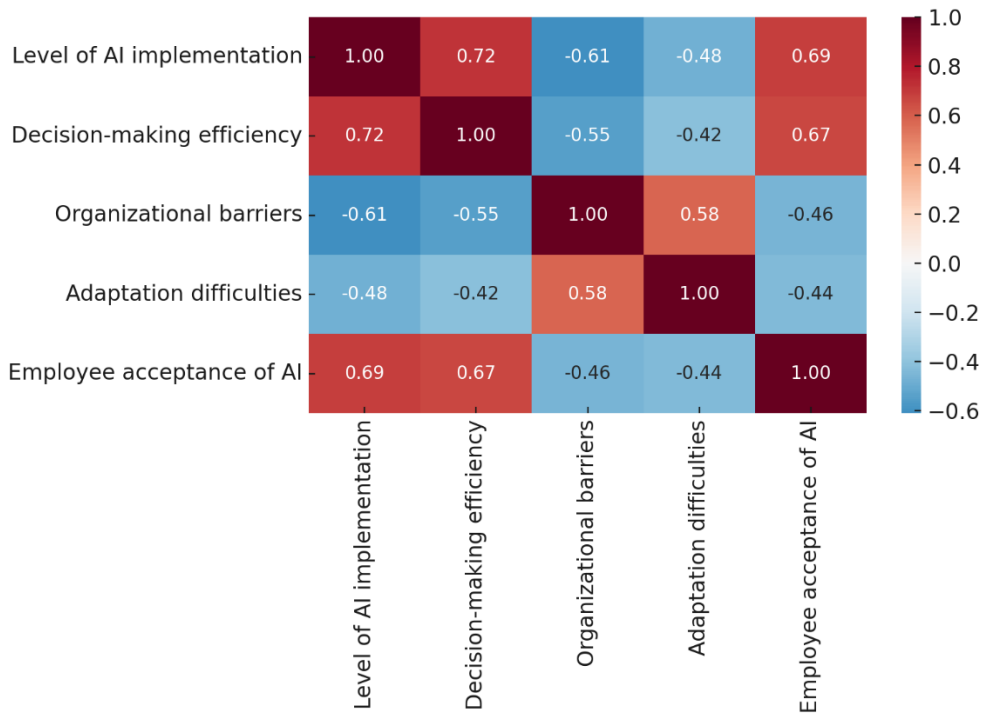

*Source: Own elaboration.*

The heatmap illustrates the strength and direction of relationships between the main variables examined in the study. The analysis shows a strong positive correlation between the level of AI implementation and both decision-making efficiency and employee acceptance of AI, indicating that organizations with higher adoption of artificial intelligence tend to experience more effective decision-making processes and greater employee openness to technological innovation.

Conversely, negative correlations appear between the level of AI implementation and organizational barriers as well as adaptation difficulties, suggesting that as AI becomes more integrated into business operations, structural and behavioral obstacles tend to decrease. The results highlight the interconnected nature of technological, organizational, and human factors in the process of AI adoption.

## 5. Conclusions and Future Research Implications

The analysis of artificial intelligence (AI) implementation in contemporary enterprises reveals a strong and multidimensional impact of this technology on decision-making processes, communication practices, and managerial effectiveness. The study demonstrates that AI is no longer perceived merely as a technological innovation but as a strategic instrument supporting organizational agility, data-driven leadership, and operational efficiency. The quantitative results based on a sample of 92 companies show that 93% of organizations have already integrated AI-based solutions, mainly in customer service (59%), data analysis and forecasting (16%), managerial decision support (15%), and process automation (10%). These findings confirm that AI has become an indispensable component of corporate strategy, particularly in enhancing decision speed, accuracy, and evidence-based reasoning.

The empirical results correspond with earlier studies by Davenport and Harris (2017) and Przegalińska and Jemielniak (2019), who emphasized the transformative influence of AI on management efficiency and strategic responsiveness. The present research expands upon their conclusions by providing quantitative evidence that AI not only improves decision-making but also fosters a cultural shift toward hybrid intelligence - where human intuition complements algorithmic analysis. Furthermore, the outcomes align with recent work by Florea and Croitoru (2025), showing that AI strengthens internal communication clarity, feedback precision, and leadership adaptability.

A key contribution of this study lies in identifying the primary barriers and competencies determining successful AI implementation. The results indicate that organizational rather than technological challenges dominate the adoption landscape: 74% of respondents agreed that implementation difficulties stem mainly from internal structures, resistance to change, and insufficient managerial competencies.

The most frequently cited barriers include employee resistance (49%), high implementation costs (25%), and legal ambiguities (24%). At the same time, the most crucial competencies for effective adoption are understanding AI mechanisms (49%) and change management (45%), while purely technical skills such as programming play a lesser role.

These findings highlight that AI-driven transformation depends as much on human and organizational factors as on technological infrastructure. Employee adaptation remains a considerable challenge, with over 90% of respondents observing difficulties in using AI tools—most often due to creating effective prompts (62%) and building trust in AI systems. However, contrary to common assumptions, job loss concerns were marginal, suggesting that AI is largely perceived as an enabler rather than a threat to employment.

Despite its contributions, the study is subject to certain limitations. Firstly, the sample size and national focus constrain the generalizability of the results; the findings may not fully reflect conditions in different cultural or economic environments. Secondly, as the research relied on self-reported data, there is a potential risk of response bias and subjective interpretation. Thirdly, the study presents a cross-sectional perspective and does not capture the long-term evolution of AI strategies or their dynamic adaptation to external changes.

Future research should therefore extend this analysis by conducting cross-sector and cross-country comparisons to identify contextual differences in AI adoption. Longitudinal studies could provide deeper insights into how AI integration evolves over time and how organizations develop resilience, innovation, and ethical maturity in response to technological change. Additionally, future inquiries could explore the interrelation between AI implementation and specific performance outcomes such as innovation capability, customer satisfaction, and leadership effectiveness.

Ultimately, the study confirms that artificial intelligence - when combined with adaptive leadership, strategic communication, and human-centered management - has the potential to redefine the foundations of modern organizational decision-making and create lasting competitive advantage.